\documentclass[aps,prl,twocolumn,showpacs,superscriptaddress,floatfix]{revtex4-1}

\usepackage{amsfonts,amsmath,graphicx}

\newcommand{\ket}[1]{|#1\rangle}
\newcommand{\Exp}[1]{\mathrm{e}^{#1}}

\begin{document}

\title{Quantum transducer in circuit optomechanics}
\author{Nicolas Didier}
\affiliation{Scuola Normale Superiore, NEST and Istituto Nanoscienze - CNR, Pisa, Italy}
\author{Stefano Pugnetti}
\affiliation{Scuola Normale Superiore, NEST and Istituto Nanoscienze - CNR, Pisa, Italy}
\author{Yaroslav M. Blanter}
\affiliation{Kavli Institute of Nanoscience, Delft University of Technology, Lorentzweg 1, 2628 CJ Delft, The Netherlands}
\author{Rosario Fazio}
\affiliation{Scuola Normale Superiore, NEST and Istituto Nanoscienze - CNR, Pisa, Italy}

\begin{abstract}
Mechanical resonators are macroscopic quantum objects with great potential.
They couple to many different quantum systems such as spins, optical photons, cold atoms, and Bose Einstein condensates.
It is however difficult to measure and manipulate the phonon state due to the tiny motion in the quantum regime.
On the other hand, microwave resonators are powerful quantum devices since arbitrary photon state can be synthesized and measured with a quantum tomography.
We show that a linear coupling, strong and controlled with a gate voltage, between the mechanical and the microwave resonators enables to create quantum phonon states, manipulate hybrid entanglement between phonons and photons and generate entanglement between two mechanical oscillators.
In circuit quantum optomechanics, the mechanical resonator acts as a quantum transducer between an auxiliary quantum system and the microwave resonator, which is used as a quantum bus.
\end{abstract}
\maketitle

Nano-mechanical systems (NMS) have been recently cooled down to their ground state~\cite{Cleland,Teufel,Painter}.
Such breakthroughs open the path toward the test of quantum mechanics for massive macroscopic systems, especially above the Planck mass $m_P\simeq22\,\mu\mathrm{g}$, where gravitation starts to play a role in the dynamics~\cite{Penrose,Aspelmeyer}.
Moreover, the displacement of a NMS can be coupled to a large variety of quantum systems, such as optical photons~\cite{Harris,Kippenberg,PainterOMC,Braive}, atoms~\cite{TreutleinSA,TreutleinBEC,TreutleinCA}, spins~\cite{Zoller,Arcizet} and even Majorana bound states~\cite{Trauzettel}. NMS naturally constitute a quantum transducer
between quantum systems of different nature, gathering their potentialities.
In particular, microwave resonators can be used to synthesize arbitrary quantum phonon states and to measure the state of the mechanical oscillator with quantum tomography~\cite{Hofheinz}.
Combining microwave photons and phonons gives rise to the field of circuit quantum optomechanics.

\begin{figure}[b]
\centering
\includegraphics[width=\columnwidth]{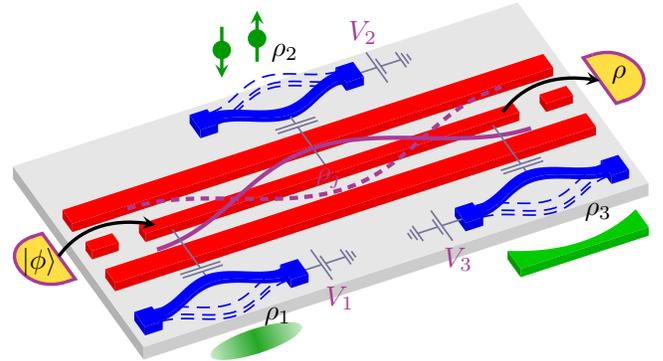}
\caption{Scheme of the system.
The mechanical resonator acts as a quantum transducer between an auxiliary quantum system (cold atoms, spins, optical photons\dots) and the microwave resonator, which is used as a quantum bus.
Arbitrary quantum state can be synthesized in the microwave resonator and measured with a quantum tomography.}
\label{figsystem}
\end{figure}

In this paper, we consider a microwave resonator coupled linearly to several mechanical oscillators.
Each mechanical resonator can be coupled to an auxiliary quantum system, as depicted in Fig.~\ref{figsystem}.
The linear coupling is obtained by adding a gate voltage on the NMS~\cite{phononblockade}.
This coupling is known in quantum optics as the ``beam splitter interaction''~\cite{zhang,tian} and gives rise to coherent oscillations between the two resonators.
The gate voltage may generate a strong coupling and allows to freeze the dynamics when the desired state is reached.
This powerful setup gives full access to the state of the mechanical oscillator, which is extremely challenging with a direct measurement.
This exciting achievement is realizable experimentally thanks to the last breakthroughs in circuit QED for the preparation and the detection of a photonic density matrix~\cite{Hofheinz}.
We demonstrate that the density matrix of the mechanical oscillator can be measured by performing quantum tomography of the microwave resonator once the phononic state has been transferred to the photons.
We show how to use this hybrid device to synthesize arbitrary quantum phonon states as well as to generate entanglement between phonons and photons.
When two mechanical oscillators are coupled to the same microwave resonator, entanglement can be created between the phonons and then swapped between one of the mechanical oscillators and the photons.

In the quantum regime, the mechanical resonator oscillates with an amplitude of the order of the zero-point fluctuations $x_\mathrm{zpf}=\sqrt{\hbar/2M\Omega}$, which is a few femtometers for a mass $M$ of few femtograms and a frequency $\Omega$ in the gigahertz range.
The mechanical oscillator and the microwave resonator are connected with the capacitance $C_g$, see Fig.~\ref{figsystem}.
When the NMS is oscillating the capacitance depends on its position $x$ and, because the motion is tiny, the interaction can be described with the linear law $C_g=C_g^{(0)}+xC_g^{(1)}$.
The ratio $C_g^{(1)}/C_g^{(0)}$ is equal to the ratio between the zero-point fluctuations and the distance between the electrodes.
This dependence gives rise to a coupling between the displacement and the electromagnetic energy of the microwave resonator, $C_g^{(1)}V^2x/2$, where $V$ is the voltage of the microwave resonator.
When the gate voltage $V_g$ is applied on the NMS, the electric field is shifted and a part of the coupling energy becomes proportional to the voltage, $C_g^{(1)}V_gVx$.
The displacement $x$ and the field $V$ can then be expressed with the phonon and photon ladder operators, respectively $a$ and $b$, according to $x=x_\mathrm{zpf}(a+a^\dag)$ and $V=V_\mathrm{rms}(b+b^\dag)$, where the root-mean-square voltage $V_\mathrm{rms}$ is of the order of microvolts.
The total coupling Hamiltonian,
$H_\mathrm{int}=\hbar g_\mathrm{rad}(a+a^\dag)b^\dag b + \hbar g(a+a^\dag)(b+b^\dag),$
is then composed of two terms:
the usual radiation pressure interaction with the strength $g_\mathrm{rad}=C_g^{(1)}V_\mathrm{rms}^2x_\mathrm{zpf}/\hbar$ and the new linear coupling with the strength $g=C_g^{(1)}V_gV_\mathrm{rms}x_\mathrm{zpf}/\hbar$~\cite{phononblockade}.
For large gate voltage $V_g\gg V_\mathrm{rms}$, the linear coupling can reach several megahertz, well above the radiation pressure interaction.
As the coupling remains far smaller than the frequency, the rotating wave approximation can be safely applied.
The total Hamiltonian of the system then reads
\begin{equation}
H=\hbar\Omega\left(a^\dag a+b^\dag b\right) + \hbar g\left(a^\dag b+b^\dag a\right) + \hbar g_\mathrm{rad}(a+a^\dag)b^\dag b.
\label{H}
\end{equation}
Here we work at the resonance between the phonons and the photons.
The beam splitter Hamiltonian can be obtained after linearizing the radiation pressure term at sufficiently strong driving~\cite{Teufelstrong}.
Hamiltonian~\eqref{H} includes the radiation pressure, but this coupling will not have a significant impact in the dynamics for $g_\mathrm{rad}<g$.

When the radiation pressure is not taken into account, the Hamiltonian conserves the total number $n=\langle a^\dag a+b^\dag b\rangle$ of excitations.
Starting from the Fock state with $n$ phonons and no photon, $\ket{n}_a\otimes\ket{0}_b\equiv\ket{n,0}$, the wave-function $\ket{\psi_n(t)}$ evolves on the basis $\{\ket{n-k,k},\;k=0,\dots,n\}$ in the form of the SU(2) coherent state~\cite{Milburn}
\begin{multline}
\ket{\psi_n(t)}=\Exp{-in\Omega t}\sum_{k=0}^n\binom{n}{k}^{1/2}(\cos gt)^{n-k}\\\times(-i\sin gt)^k\,\ket{n-k,k},
\end{multline}
rotating around the Bloch sphere with the period $T=2\pi/g$.
Two times are very interesting during the temporal evolution, the quarter-period $T_{1/4}=\pi/2g$ and the eighth-period $T_{1/8}=\pi/4g$.
Indeed, $\ket{\psi_n(T_{1/4})}=\Exp{-in\theta}\ket{0,n}$, which means that the phononic Fock state is transferred into the corresponding photonic Fock state after a quarter of period, up to the phase $-n\theta$, where $\theta=(1+\Omega/g)\pi/2$, i.e.,
\begin{equation}
\ket{\psi(0)}=\ket{\phi,0}\ \Rightarrow\ \ket{\psi(\pi/2g)}=\ket{0,\Exp{-i\hat{n}\theta}\phi}.
\end{equation}
The phase $\theta$ accumulated can be bypassed by tuning the ratio $\Omega/g$ with the gate voltage.
The accuracy of the transferred state can be quantified with the fidelity $F(t)=|\langle0,\phi\ket{\psi(t)}|$ of the pure state.
The fidelity turns out to be equal to $F(t)=\sum_n|\langle\phi\ket{n}|^2|\sin gt|^n$, which is equal to unity at $t=T_{1/4}$.
The synthesis is thus perfect, and the phonon state can be measured from quantum tomography of the microwave resonator.
Inversely, if one desires to have the phonon state $\ket{\phi}_a$, the photon state $\ket{\Exp{i\hat{n}\theta}\phi}_b$ has to be synthesized in the microwave resonator.
The possibility to synthesize an arbitrary quantum state for microwave photons, demonstrated in the experiment of Ref.~\onlinecite{Hofheinz}, enables to generate an arbitrary quantum phonon state.
Once the appropriate photonic state is prepared, the coupling is switched on during a time $T_{1/4}$, and the desired phononic state is obtained with a maximum fidelity.
For instance, a superposition of various coherent phononic states $\ket{\alpha}_a$ is obtained from the coherent photonic states $\ket{\alpha+\theta}_b$.
When the linear coupling is switched off, the radiation pressure interaction is still present and can in principle affect the transferred state.
We however checked numerically that the fidelity is almost not affected for $g_\mathrm{rad}<g$.

\begin{figure}[t]
\centering
\includegraphics[height=6cm]{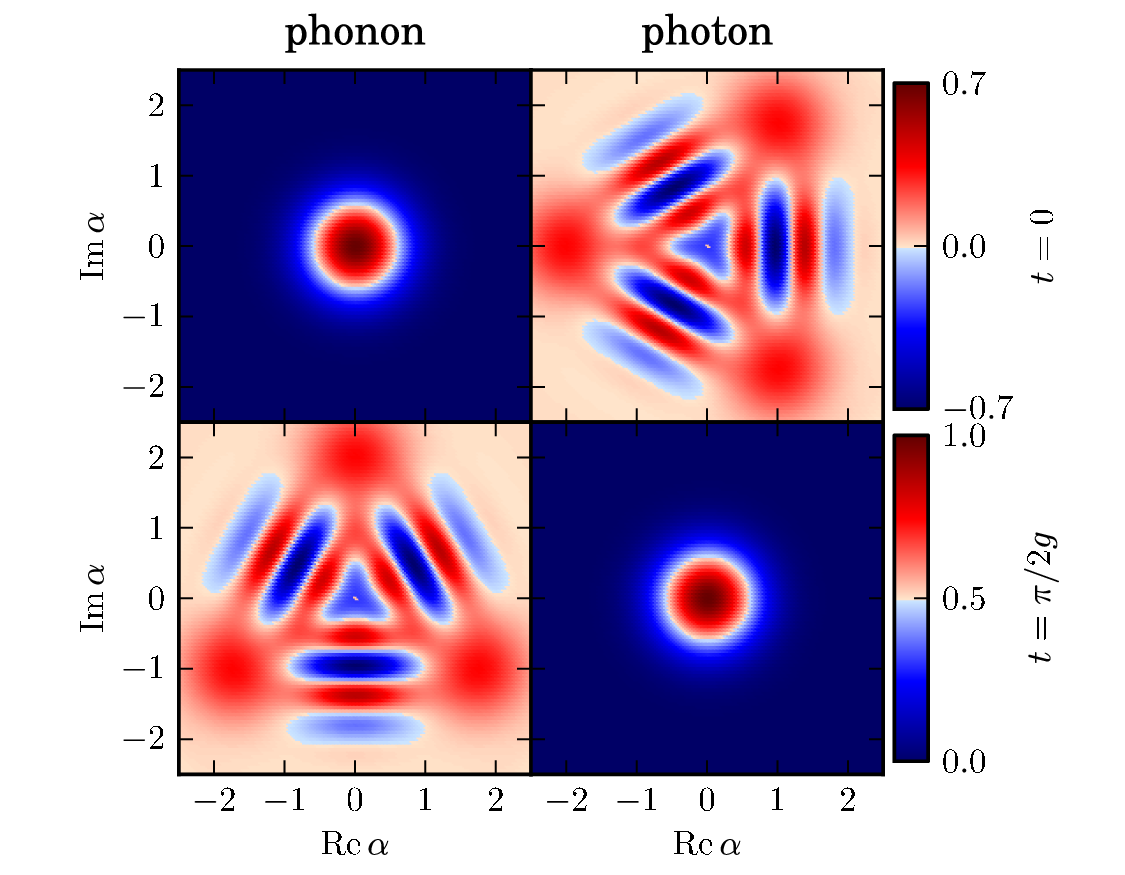}
\caption{(Color online) 
Wigner function $W(\alpha)$ of the voodoo cat state.
The left plots concern the phonons and the right ones the photons.
After a time $T_{1/4}=\pi/2g$, the initially prepared photonic state is transferred to the phonons with a fidelity $F=99.7\,\%$.}
\label{figvoodoo}
\end{figure}

After an eighth of period, the state is a macroscopic superposition of states $\ket{k,n-k}$
\begin{equation}
\ket{\psi_n(T_{1/8})}=\frac{\Exp{-in\pi\Omega/4g}}{\sqrt{2}^n}\sum_{k=0}^n(-i)^k\binom{n}{k}^{1/2}\,\ket{k,n-k}.
\label{generalentangled}
\end{equation}
In the case where one photon is initially generated, $\Exp{i\pi\Omega/4g}\ket{0,1}$, the maximally entangled Bell state $\ket{\Psi}$ is created
\begin{equation}
\ket{\Psi}=\frac{1}{\sqrt{2}}\left[\ket{0,1}-i\ket{1,0}\right].
\label{bellstate}
\end{equation}
The entanglement of this pure state can be estimated with the von Neumann entropy $S_V$~\cite{Vedral,Fazio} of the microwave resonator, defined from its density matrix $\rho_b$,
\begin{equation}
S_V=-\mathrm{Tr}\left\{\rho_b\log_2\rho_b\right\}.
\end{equation}
The von Neumann entropy is equal to unity for the Bell state $\ket{\Psi}$ and can be directly calculated from the quantum tomography measurement of the microwave resonator.
This experimental technique has been recently developed in the experiments of Refs.~\onlinecite{Wallraff_tomo, Lehnert} for itinerant microwave photons.

\begin{figure}[t]
\centering
\includegraphics[height=6cm]{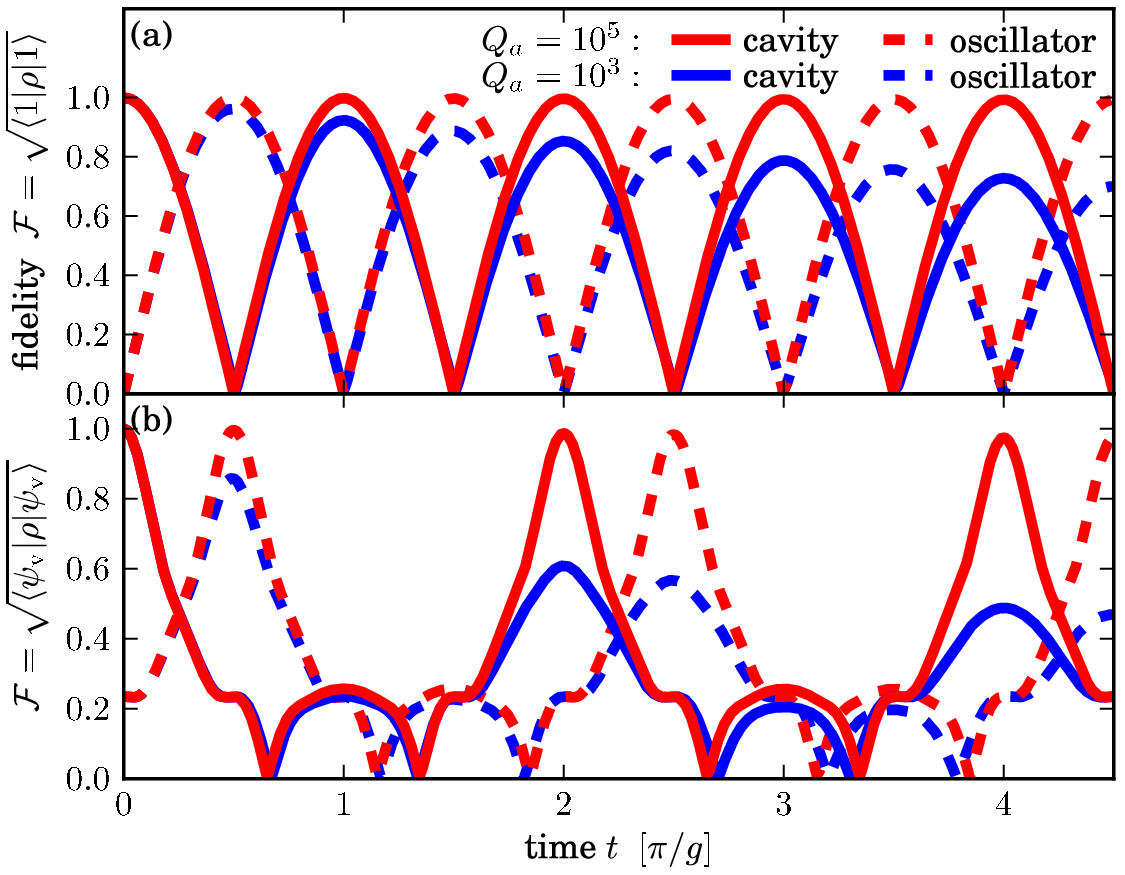}
\caption{(Color online) 
Rabi oscillations for different values of the quality factor $Q_a$ of the mechanical oscillator.
The initial state is the Fock state $\ket{0}_a\otimes\ket{1}_b$ in (a) and the voodoo cat state $\ket{0}_a\otimes\ket{\psi_\mathrm{v}}_b$ in (b).
The state is transferred to the mechanical oscillator at time $T_{1/4}=\pi/2g$ (${}_a\langle\phi|\rho|\phi\rangle_a$, dashed lines) and recovered in the microwave resonator after a period $T$ (${}_b\langle\phi|\rho|\phi\rangle_b$, full lines).}
\label{figrabi}
\end{figure}

\begin{figure}[t]
\centering
\includegraphics[height=6cm]{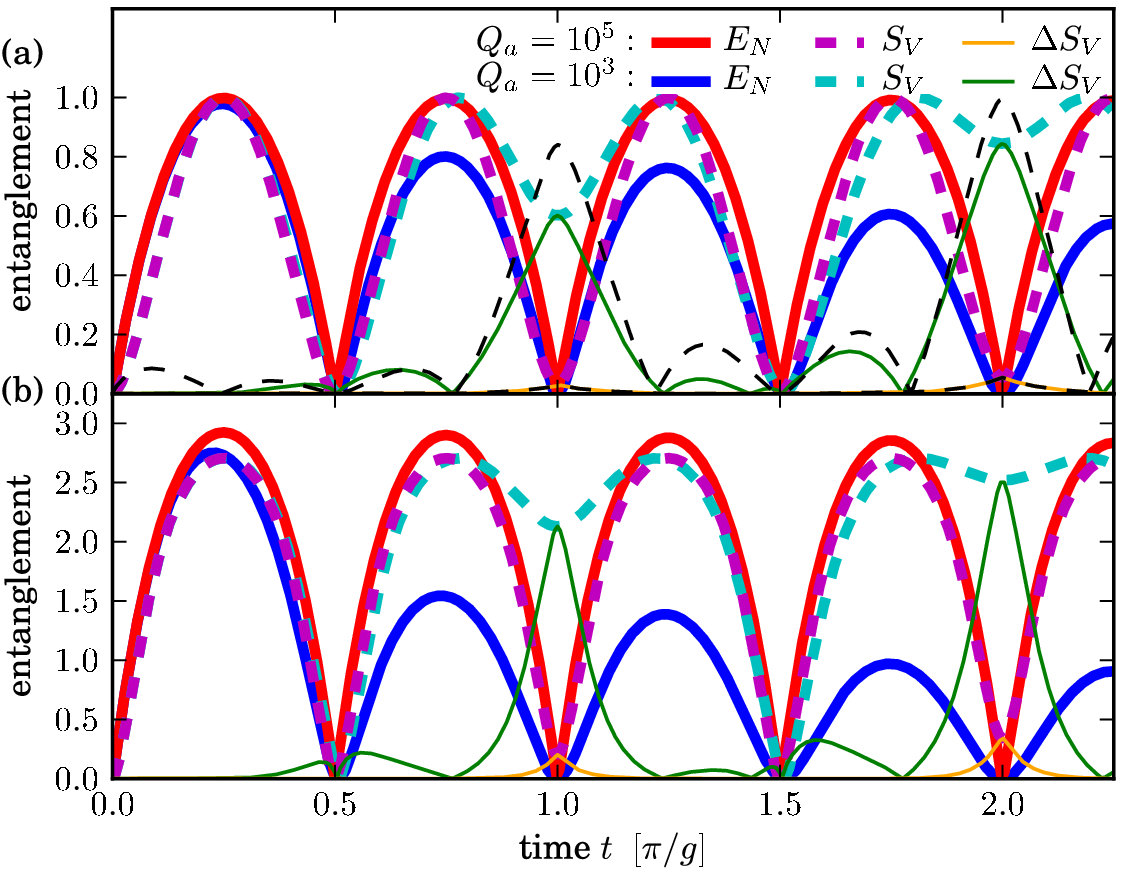}
\caption{(Color online) 
Temporal evolution of the entanglement.
The initial state comprises one photon, $\ket{0}_a\otimes\ket{1}_b$, in (a) and 10 photons, $\ket{0}_a\otimes\ket{10}_b$, in (b).
The logarithmic negativity is compared to the von Neumann entropy of the cavity.
The good agreement of these two quantities when the oscillations are coherent shows that the measurement of the von Neumann entropy is a good indication of entanglement.
The maximal entanglement is found at the time $T_{1/8}=\pi/4g$, equal to $E_N\simeq1$ in (a) and $E_N=0.85\log_2(11)$ in (b).
The contribution of dissipation in the von Neumann entropy $\Delta S_V$ is maximum at $T_{1/2}$ and $T$.
The thin dashed black lines in (a) are the results Eq.~\eqref{eqtoy}.}
\label{figentangle}
\end{figure}

The presence of dissipation due to the unavoidable coupling to the environment changes the dynamics of the system.
The finite lifetime of the phonons and the photons, inverse of the damping rates $\gamma_a$ and $\gamma_b$, can be taken into account with the introduction of Lindbladian operators~\cite{lindbladian}
\begin{equation}
L_a\rho=\tfrac{1}{2}\gamma_a\left(2a\rho a^\dag-a^\dag a\rho-\rho a^\dag a\right),
\end{equation}
and similarly $L_b$ for the cavity.
The quantum dynamics of the total density matrix $\rho$ is then governed by the Hamiltonian $H$ and the Lindbladian according to the quantum master equation,
\begin{equation}
\partial_t\rho(t)=\frac{1}{i\hbar}\left[H,\rho(t)\right] + (L_a+L_b)\rho(t).
\end{equation}
The quality factors $Q_{a,b}=\Omega/\gamma_{a,b}$ of the resonators is chosen to be equal to $10^5$ in the numerical calculations.
The frequency is set to $\Omega/2\pi=1\,\mathrm{GHz}$, the linear coupling $g/2\pi=10\,\mathrm{MHz}$ ($\theta=\pi/2$), and the radiation pressure coupling $g_\mathrm{rad}=g/100$.
We define the fidelity of the target state $\ket{\phi}$ as follows
\begin{equation}
F=\sqrt{\langle\phi|\rho\ket{\phi}}.
\end{equation}
The state can be visualized with the Wigner function $W(\alpha)$ of the density matrix,
\begin{equation}
W_{j=a,b}(\alpha)=\mathrm{Tr}_j\left\{D_j^\dag(\alpha)\rho_jD_j(\alpha)\Pi\right\},
\end{equation}
where $\rho_j$ is the reduced density matrix of one subsystem, the operator $D_{a,b}(\alpha)$ is the displacement operator $D_a(\alpha)=\exp(\alpha a^\dag-\alpha^*a)$, similar for $D_b$, and $\Pi$ is the parity operator.
The quantum tomography is experimentally performed by displacing the field with different microwave drive pulses to build the Wigner function of the state~\cite{Hofheinz}.
For illustration, the synthesis of the phononic voodoo cat state $\ket{\psi_\mathrm{v}}=\left[\ket{2i}+\ket{2i\Exp{i2\pi/3}}+\ket{2i\Exp{-i2\pi/3}}\right]/\sqrt{3}$ is presented in Fig.~\ref{figvoodoo}.
The resulting state of the mechanical oscillator is the superposition of three Gaussian states centered at $x=0,\,\pm\sqrt{3}x_\mathrm{zpf}$, the ``alive'', ``dead'', and ``zombie'' states.
The quantum coherence of the coupled system can be probed with a Rabi spectroscopy, as presented in Fig.~\ref{figrabi}.
The state of the microwave resonator is measured continuously during the dynamics and compared to the initial state.
Quantum revivals appear with the period $T$ and indicate that the interaction is quantum coherent.
The voodoo cat state is obtained with a fidelity of $99.7\,\%$ and the Fock state with 1 photon is transferred with a fidelity of $99.9\,\%$.

The appropriate measure of entanglement of a bipartite system in the presence of dissipation is the logarithmic negativity~\cite{Vidal,Fazio}, defined as 
\begin{equation}
E_N=\log_2\|\rho^{\mathrm{T}_a}\|,
\end{equation}
where $\|\rho\|=\mathrm{Tr}\sqrt{\rho^\dag\rho}$ and $\rho^{\mathrm{T}_a}$ is the partial transpose with respect to the mechanical oscillator.
Starting from a photonic state, the entanglement is maximum at the time $T_{1/8}$, then vanishes at the time $T_{1/4}$ and so on with a period $T_{1/4}$.
The measurement of the logarithmic negativity needs a quantum tomography of the whole system, which is not possible on the mechanical oscillator.
However we show that the von Neumann entropy, accessible experimentally in the microwave resonator, is a good indication of entanglement in the strong coupling regime $g\gg\gamma_a$.
Indeed, in Fig.~\ref{figentangle} presenting the temporal evolution of the entanglement, the logarithmic negativity and the von Neumann entropy have very close dynamics for sufficiently high quality factors.
To quantify the contribution to the von Neumann entropy from the incoherent mixing induced by dissipation, let us define the quantity
\begin{equation}
\Delta S_V=\left| \left. S_V\right|_{\gamma_a,\gamma_b} -\left. S_V\right|_{\gamma_a=\gamma_b=0} \right|.
\end{equation}
In the case where one excitation is involved, $\Delta S_V$ can be evaluated with a simple model~\cite{note}.
The part of the entropy due to dissipation, $\Delta S_{V,n=1}$, has maxima at times $t\in\{T_{1/2}, T,\dots\}$ and vanishes at $\{T_{1/4}, 3T_{1/4},\dots\}$ and close to $\{T_{1/8}, 3T_{1/8},\dots\}$.
Moreover, at times multiple of the period $T$, the initial state is fully transferred to one of the two resonators and the entanglement vanishes, i.e. $\Delta S_V(T)=S_V(T)$.
As a consequence, measuring a vanishing value of $S_V$ at time $T$ indicates that the contribution due to dissipation in the von Neumann entropy measured at time $T_{1/8}$ is negligible,
\begin{equation}
S_V(T)\ll1\ \Rightarrow\ S_V(T_{1/8})\equiv E_N(T_{1/8}),
\end{equation}
which can be used to prove the hybrid entanglement between phonons and photons.

When two mechanical oscillators are coupled to the microwave resonator, an entangled state equivalent to Eq.~\eqref{generalentangled} is obtained between the two oscillators after a time $T_0=\pi/2\sqrt{2}g$ from the initial Fock state $\ket{n}_b\otimes\ket{0,0}_a$ of the cavity~\cite{Haroche} and can be detected by entanglement swapping.
As an example, starting with one photon, the Bell state Eq.~\eqref{bellstate} between the oscillators is created after the time $T_0$.
To detect entanglement, one of the two linear couplings can then be switched off during the time $T_{1/4}$ to transfer the Bell state of the oscillators to the hybrid Bell state Eq.~\eqref{bellstate} between one oscillator and the cavity.
The method described in the previous paragraph can be applied to detect entanglement from the cavity,
\begin{align}
&\ket{\psi(0)}=\ket{1}_b\otimes\ket{0,0}_a && \nonumber\\
\Rightarrow&\ket{\psi(T_0)}=\ket{0}_b\otimes\left[\ket{0,1}_a+\ket{1,0}_a\right] && (g_1=g_2=g) \nonumber\\
\Rightarrow&\ket{\psi(T_1)}=\ket{0}_{a_2}\otimes\left[\ket{0,1}-i\ket{1,0}\right] && (g_1=0)
\end{align}
where global phases and normalizations are not explicitly written, $g_{1,2}$ refer to the linear couplings and $T_1=T_0+T_{1/4}$.
If the two oscillators were not entangled, a vanishing von Neumann entropy would be measured on the cavity after the time $T_1$.

The highly accurate density matrix transfer between the mechanical resonator and the microwave resonator together with the possibility to synthesize arbitrary photonic state and to perform quantum tomography on the microwave resonator can be applied to other quantum systems.
As depicted in Fig.~\ref{figsystem}, we consider an architecture composed of several quantum transducers (the mechanical oscillators) coupled to a quantum bus (the microwave resonator).
Each quantum transducer is coupled to an auxiliary quantum system.
As an example, the coupling of a NMS to a spin is of the Jaynes Cumming type~\cite{Zoller,Arcizet,TreutleinBEC}, the coupling to a cloud of cold atoms is linear~\cite{TreutleinSA,TreutleinCA}, and the coupling to optical photons is mediated with the radiation pressure force~\cite{Harris,Kippenberg}.
Information can thus be transferred between the NMS and the auxiliary quantum system,
the individual couplings being controlled by the gate voltages.
The information about the auxiliary quantum systems is communicated to the mechanical resonator, whose density matrix is then transferred to the microwave resonator and measured with a quantum tomography.
Inversely, a quantum state can be synthesized in the microwave resonator, transferred to the mechanical resonator and communicated to the auxiliary quantum system.
Communications between two auxiliary quantum systems can be also transferred directly through the transducers and the bus.

In conclusion, the linear coupling of a mechanical oscillator and a microwave resonator is a powerful device to measure and synthesize quantum phonon states and to generate and detect entanglement between phonons and photons.
The entanglement of two mechanical oscillators can be generated and detected by the cavity after entanglement swapping.
Circuit quantum optomechanics opens the way to a rich physics where the intrinsic properties of electrons, photons, and phonons can be exploited.
NMS also enable the integration of quantum systems such as cold atoms, spins and optical photons in circuit quantum optomechanics.

The authors thank Giulia~Ferrini, Vittorio~Giovannetti, Gordey~Lesovik, Pierre~Meystre, Mika~Sillanp\"a\"a, and Heung-Sun~Sim for useful questions and discussions.
We acknowledge financial support from EU through the projects QNEMS, SOLID and GEOMDISS.


\begin{thebibliography}{99}
%
\bibitem{Cleland}
	A.~D.~O'Connell \textit{et al.},
	Nature \textbf{464}, 697 (2010).
\bibitem{Teufel}
	J.~D.~Teufel \textit{et al.},
	Nature \textbf{475}, 359 (2011).
\bibitem{Painter}
	J.~Chan \textit{et al.},
	Nature \textbf{478}, 89 (2011).
\bibitem{Penrose}
	W.~Marshall, C.~Simon, R.~Penrose, and D.~Bouwmeester
	Phys.~Rev.~Lett. \textbf{91}, 130401 (2003).
\bibitem{Aspelmeyer}
	O.~Romero-Isart \textit{et al.},
	Phys.~Rev.~Lett. \textbf{107}, 020405 (2011).
\bibitem{Harris}
	J.~D.~Thompson \textit{et al.},
	Nature \textbf{462}, 72 (2008).
\bibitem{Kippenberg}
	G.~Anetsberger \textit{et al.}, 
	Nature~Phys. \textbf{5}, 909 (2009).
\bibitem{PainterOMC}
	M.~Eichenfield \textit{et al.},
	Nature \textbf{462}, 78 (2009).
\bibitem{Braive}
	E.~Gavartin \textit{et al.}, 
	Phys.~Rev.~Lett. \textbf{106}, 203902 (2011).
\bibitem{TreutleinSA}
	K.~Hammerer \textit{et al.},
	Phys.~Rev.~Lett. \textbf{103}, 063005 (2009).
\bibitem{TreutleinBEC}
	D.~Hunger \textit{et al.},
	Phys.~Rev.~Lett. \textbf{104}, 143002 (2010).
\bibitem{TreutleinCA}
	S.~Camerer \textit{et al.},
	Phys.~Rev.~Lett. \textbf{107}, 223001 (2011).
\bibitem{Zoller}
	P.~Rabl \textit{et al.},
	Nat.~Phys. \textbf{6}, 602 (2010).
\bibitem{Arcizet}
	O.~Arcizet \textit{et al.},
	Nature~Phys. \textbf{7}, 879 (2011).
\bibitem{Trauzettel}
	S.~Walter, T.~L.~Schmidt, K.~B\o rkje, and B.~Trauzettel,
	Phys.~Rev.~B \textbf{84}, 224510 (2011).
\bibitem{Hofheinz}
	M.~Hofheinz \textit{et al.},
	Nature \textbf{459}, 546 (2009).
\bibitem{phononblockade}
	N.~Didier, S.~Pugnetti, Ya.~M.~Blanter, and R.~Fazio,
	Phys. Rev. B \textbf{84}, 054503 (2011).
\bibitem{zhang}
	J. Zhang, K. Peng, and S. L. Braunstein,
	Phys. Rev. A \textbf{68}, 013808 (2003).
\bibitem{tian}
	L. Tian, and H. Wang,
	Phys. Rev. A \textbf{82}, 053806 (2010).
\bibitem{Teufelstrong}
	J.~D.~Teufel \textit{et al.},
	Nature \textbf{471}, 204 (2011).
\bibitem{Milburn}
	G. J. Milburn, J. Corney, E. M. Wright, and D. F. Walls,
	Phys.~Rev.~A \textbf{55}, 4318 (1997).
\bibitem{Vedral}
	V.~Vedral, M.~B.~Plenio, M.~A.~Rippin, and P.~L.~Knight,
	Phys.~Rev.~Lett. \textbf{78}, 2275 (1997).
\bibitem{Fazio}
	L.~Amico, R.~Fazio, A.~Osterloh, and V.~Vedral,
	Rev.~Mod.~Phys. \textbf{80}, 517 (2008).
\bibitem{Wallraff_tomo}
	C.~Eichler \textit{et al.},
	Phys.~Rev.~Lett. \textbf{106}, 220503 (2011).
\bibitem{Lehnert}
	F.~Mallet \textit{et al.},
	Phys.~Rev.~Lett. \textbf{106}, 220502 (2011).
\bibitem{lindbladian}
	M.~O.~Scully and M.~S.~Zubairy, \textit{Quantum Optics}, Cambridge university press (1997).
\bibitem{Vidal}
	G.~Vidal and R.~F.~Werner,
	Phys.~Rev.~A \textbf{65}, 032314 (2002).
%
\bibitem{note}
The result is obtained with a simple model involving the states $\ket{0,0}$, $\ket{0,1}$, $\ket{1,0}$ and $\ket{1,1}$, resulting in
\begin{multline}
\Delta S_{V,n=1}\simeq\left| \frac{\gamma_at}{\ln2} -\sin^2gt\,\log_2\!\left(1+\frac{\Exp{\gamma_at}-1}{\sin^2gt}\right) \right.\\\left. -\cos^2gt\,(1-\Exp{-\gamma_at})\log_2\!\left(\tan^2gt+\frac{\Exp{\gamma_at}-1}{\cos^2gt}\right)\right|.
\label{eqtoy}
\end{multline}
%
\bibitem{Haroche}
	A.~Rauschenbeutel \textit{et al.},
	Science \textbf{288}, 2024 (2000).
%
\end{thebibliography}
\end{document}